# Sparse identification of nonlinear dynamics applied to the levitation of acoustically large objects

Mehdi Akbarzadeh[1], Ben Halkon, and Sebastian Oberst

*Centre for Audio, Acoustics and Vibration, University of Technology Sydney, Sydney, Australia.*

## ABSTRACT

Many studies on acoustic radiation forces focus on characterizing the behavior of acoustic fields. However, the dynamic response of objects, particularly those larger than the wavelength, remains underexplored. Here we bridge this gap by deriving nonlinear equations of motion for a trapped spherical object under acoustic radiation forces and external excitation, where the Gorkov formulation fails to provide accurate results. Using Sparse Identification of Nonlinear Dynamical Systems (SINDy), we derive the corresponding nonlinear equation of motion from analytical time series data obtained through the Gorkov formulation and external excitation for acoustically small objects, and which recovers the governing equation with less than 0.05% error in coefficient values compared to the analytical solution.. We conduct experiments with the TinyLev levitator with external excitation to generate time series for acoustically large particles. Then, SINDy is applied to reconstruct governing equations from experimental data to see how external excitation amplitude influences the dynamics of acoustically large objects. Our findings demonstrate that the SINDy can effectively be used as a tool for deriving governing equations from complex data to improve and refine theoretical developments; in the present case, for acoustically large objects, where the Gorkov formulation fails to provide an accurate prediction.

## Introduction

Acoustic radiation forces emerges when incident acoustic waves interact with waves scattered from an object [1-5]. These forces have many applications and are utilized for drug delivery [6], non-contact excitation [7-9], particle manipulation [10], or vibroacoustic spectrography of imaging acoustic responses of internal body tissues to mechanical excitation [11]. The acoustic radiation force is mathematically derived by integrating the acoustic radiation stress tensor over an object's surface. The stress tensor is obtained by applying perturbations to the Navier-Stokes equations, which result in non-zero second order terms. The time-averaging method cancels out the linear, first-order term due to symmetrical oscillation, leading to a zero contribution over time. However, the nonlinear, second order terms do not cancel out and produce the non-zero, time-averaged acoustic radiation force [12-14].

---

[1] Corresponding Authors: Mehdi.Akbarzadeh@uts.edu.au

Gorkov used the time-averaged potential and kinetic energies of an acoustic field, conceptualized as the gradient of a potential function determines the density, compressibility, and the geometry as functions of the acoustic radiation force, which pushes the object toward the minimum potential. However, this formulation treats the object as an ideal particle and assumes it is acoustically small—i.e., its dimensions are much smaller than the wavelength of the incident field. As a result, Gorkov's theory fails to capture the more complex dynamical behavior exhibited by acoustically large or deformable objects [15]. While several studies have investigated the acoustic waves-object interactions for object sizes larger than the wavelength limit [16-20] a more research is required to understand their dynamics better.

The Sparse Identification of Nonlinear Dynamical (SINDy) algorithm is a data-driven method that uses optimization algorithms to construct interpretable models by identifying a sparse set of candidate functions that describe the system's behavior [21-29]. SINDy reconstructs governing equations directly from time series data by using a sparse set of predefined functions, which often effectively capture the dominant nonlinearities. We collect theoretical time series data for acoustically small objects using the Gorkov formulation and compare this with experimental time series data generated from acoustically large objects and compare their bifurcation diagrams. Then we use a statistical analysis to identify the average quantities of coefficients used for SINDy to study acoustically large objects using the extracted equations with the aim to offer a framework to better understand the dynamics of arbitrary levitated objects, beyond the limitations of classical particle-based theories

## Methodology and extracting time series data

Next, we outline the process of generating and extracting theoretical and experimental time series data; additional details can be found in the Supplementary material (S1 and S2).

### Extracting time series data using the Gorkov formulation

The Gorkov theory is valid for analyzing acoustically small objects ($kr < 1$) with $k$ and $r$ representing the acoustic wave number and the object's radius [15]. By externally exciting an acoustic radiation field periodically using e.g. an electrodynamic shaker with amplitude of $A_{\text{ex}}$, and frequency of $\omega_{\text{ex}}$, the levitated object vibrations at amplitude $A_{\text{o}}$, and frequency $\omega_{\text{o}}$ [30-36] and the resulting acoustic trap behaves as a Duffing-like oscillator [35]

$$\ddot{\theta} + C_1|\dot{\theta}|\theta + C_2(\theta - \theta^3/6) - C_3 \cos(\theta)\sin(\omega_o t) = 0, \quad (1)$$

when $\theta = \gamma_2 z$ is the non-dimensional distance $z$ of the particle, from its equilibrium point. The remaining coefficients in Eq. (1) can be defined in terms of the properties of the object and the surrounding medium.

**Extracting time series data using experimental set up**

To measure the dynamics of an object trapped in an acoustic radiation force field ($kr > 1$) we setup an experiment using a Tiny-Lev levitator [33] (Fig. 1a). The levitator is driven by a square-wave having a peak-to-peak voltage of 9 V (function generator MFG-2260MFA), and is mounted vertically to an electrodynamic shaker (Bruel & Kjaer: LDS V201) driven by an amplifier (Bruel & Kjaer: LDS LPA100) using a sinusoidal signal of amplitude between 0 mm and 3 mm and a frequency range of 0 Hz to 250 Hz. The levitated object's vibration was measured using a tripod-mounted portable digital Laser Doppler Vibrometer (LDV, Polytec: PDV-100, sensitivity factor ($velo$: 125 mm/s/volt). The levitated object was measured through a 4 mm hole at the center of the levitator's head plate (Fig.1b). The excitation and the LDV signal were monitored using an oscilloscope (Keysight: InfiniiVision DSOX2004a) recorded at 4 kHz sampling frequency. Protractors integrated into the tripod structures facilitated angular adjustments with a resolution of $\pm 0.5°$ (Fig.1c).

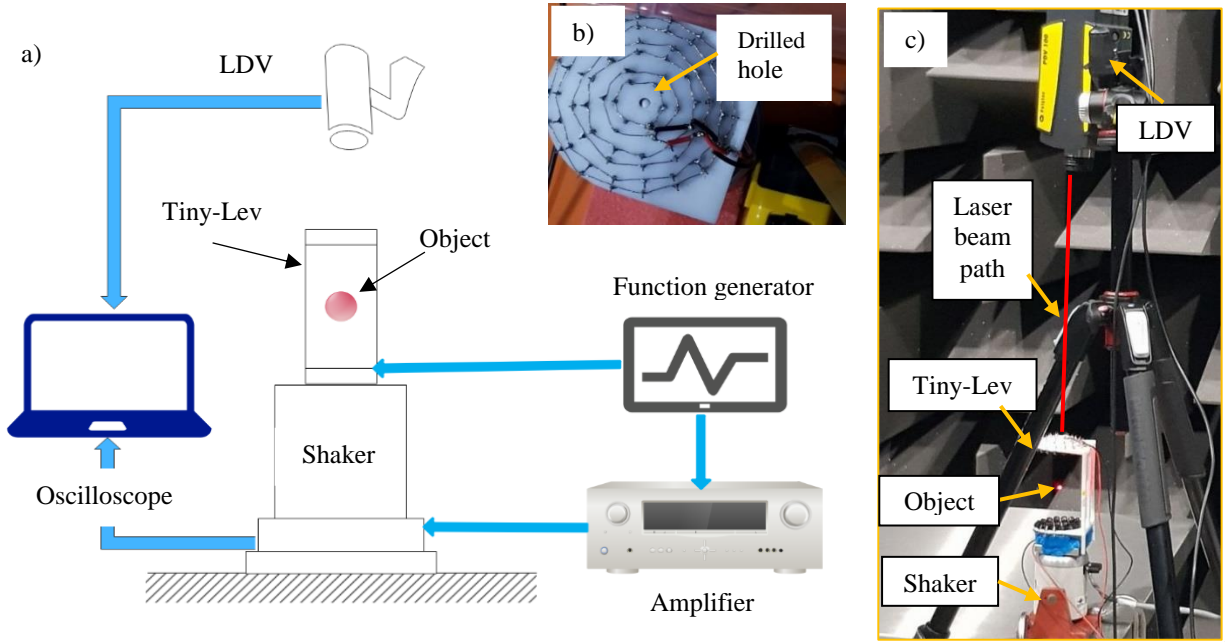

**Fig.1.** Experimental setup in hemi-anechoic chamber. a) Schematic and b) drilled hole in the Tiny-Lev's top plate showing the object, and c) Photograph of a levitated object within the Tiny-Lev mounted on the vibrating shaker, using an LDV for measurements.

Our experiments were conducted in a hemi-anechoic chamber at UTS, designed to minimize the effects of environmental noise and perturbations. The background noise was measured at the center of the chamber following ISO 26101 guidelines using a Brüel & Kjær microphone Type 4191 and a sound level meter Type 2250. The measurement includes both airborne and structure-borne noise and accounts for electrical noise in the instrumentation. The averaged background noise level over the one-third-octave bands centered between 12.5 Hz and 20 kHz was found to be 17.1 dBA, which is sufficiently low to permit accurate and precise acoustic measurements. The noise floor measured with the Type 2250 is considered reliable, as it has much lower electrical self-noise compared to the microphone. The properties of the levitated object and the host fluid were assumed to be $\rho_0 = 1.19 \frac{\text{kg}}{\text{m}^3}$ for the density of air, and for the speed of sound $c_0 = 340 \frac{\text{m}}{\text{s}}$; while the drag coefficient was set to $C_\text{d} = 1.06$ [35,36]. We used an Expanded Polystyrene (EPS) ball (radius $r = 1.87$ mm) with speed of sound of $c_1 = 540 \frac{\text{m}}{\text{s}}$, and a density of $\rho_\text{p} = 25 \frac{\text{kg}}{\text{m}^3}$. The EPS ball was levitated in the middle between the two planes of the Tiny-Lev and its vibration velocity amplitude $V_o$ had been estimated over voltage $v_o$, measured at the LDV and its sensitivity factor (*velo*) using

$$V_o = \frac{velo}{4} \times v_o. \tag{2}$$

By assuming a sinusoidal motion its amplitude of displacement ($A_\text{o}$) is determined as

$$A_\text{o} = V_\text{o}/\omega_\text{o}. \tag{3}$$

We carefully applied the GHKSS filter for noise reduction [37-40] to the extracted and converted time series to preserve the dynamics of the nonlinear system as much as possible (details Supplementary material, S3).

## Results and discussion

We apply the SINDy algorithm to analyze the dynamics of acoustically levitated objects. First, we validate its ability to reconstruct the original differential equation from time series generated by an analytical model based on Gorkov's formulation for acoustically small objects. We then extend the study to acoustically large objects, applying the analytical model and using SINDy to determine the governing equation. We implemented the SINDy algorithm in MATLAB using the Thresholded Least Squares Algorithm (TLSA) as originally introduced by Brunton et al. [21]. This algorithm iteratively applies ordinary least squares (OLS) regression, followed by thresholding small coefficients to zero, thus promoting model sparsity. We used a fixed sparsity threshold of $\lambda = 10^{-6}$ across all experiments and simulations, which provided reliable results for the identification of governing equations. The choice of $\lambda$ was based on trial-and-error tuning to achieve a balance between model accuracy and interpretability.

## SINDy for acoustically small object: Theoretical model

To benchmark the SINDy algorithm for the theoretical study described in Eq. (1) and by using the Taylor series expansion for $\sin(\theta)$ up to order 3, we apply SINDy for an acoustically small object. Training data is collected over a time interval of two seconds, with a time step size of of $dt \approx \Delta t = 10^{-6}$ s to construct state space vectors as in the following

$$\frac{d}{dt} x_1 = x_2 = \dot{\theta}, \text{ and} \tag{4}$$

$$\frac{d}{dt} x_2 = \ddot{\theta} = -C_1 |x_2| x_2 - C_2 x_1 + \frac{C_2}{6} x_1^3 + C_3 \cos(x_1) A_o \sin(\omega_o t) \tag{5}$$

to find the unknown coefficients $C_i$, ($i = 1,2,3$). A library of candidate functions is constructed, consisting of trigonometric and polynomial functions up to the fourth order using

$$\mathbf{\Theta}(\mathbf{X}) = [x_1, x_2, x_1 x_2, \dots, x_1^2, \dots, x_2^4 \text{ }], \text{ and} \tag{6}$$

$$\mathbf{u}(\mathbf{X}; t) = [x_1, x_2, \sin x_1 \sin x_2, \cos x_1, \cos x_2 \text{ }] \sin(\omega_o t).$$

Here, $\mathbf{X}(t) = [x_1, x_2]$ denotes the state vectors, and $\mathbf{\Theta}(\mathbf{X})$ is a library matrix of potential candidate functions to estimate the equation of motion, and $\mathbf{u}(\mathbf{X}; t)$ is a library matrix of potential candidate functions to estimate external excitation part. The following parameters are used as properties of the object and the acoustic medium: the sound wave frequency of the acoustic field assumed to be $f = 40$ kHz, and the acoustic medium had a density of $\rho_0 = 1.19 \ \frac{\text{kg}}{\text{m}^3}$, a speed of sound of $c_0 = 340 \ \frac{\text{m}}{\text{s}}$, and a drag coefficient of $C_d = 1.06$; the objects radius was $r = 0.71$ mm, and its material had a density of $\rho_1 = 25 \ \frac{\text{kg}}{\text{m}^3}$ and speed of sound of $c_1 = 570 \ \frac{\text{m}}{\text{s}}$. Using a fourth-order Runge-Kutta method on Eq. (1) we generate three benchmark systems of different dynamical responses to evaluate the performance of SINDy, using the amplitudes $\mathbf{A_{in}} = [0.15, 0.25, \text{and } 0.25]$ mm and excitation frequencies $\boldsymbol{\omega_o} = [34, 600, 600]$ Hz based on bifurcation diagram of the analytical solution [41]. Results are shown in Table 1.

**Table 1.** Application of the SINDy algorithm to three theoretical benchmark (B) models based on the Gorkov formulation, including extracted coefficients (Coeff.) and phase portraits. Minimal deviations from the theoretical results are observed using an acoustically small object.

| | Coeff. | Theory | SINDy | Phase space (Theory) | Phase space (SINDy) |
|---|---|---|---|---|---|
| B1 | $C_1$ | 0.0149 | 0.0149 | 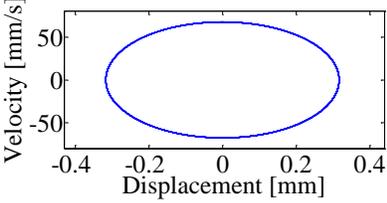 | 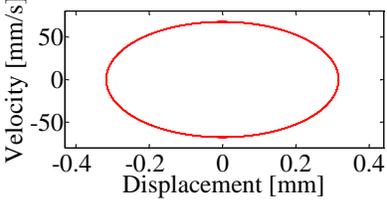 |
| | $C_2$ | 78,941 | 78,943 | | |
| | $C_3$ | 15,571 | 15,578 | | |
| B2 | $C_1$ | 0.0149 | 0.0149 | 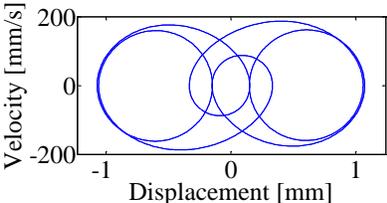 | 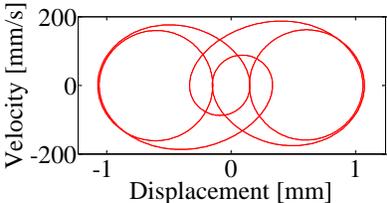 |
| | $C_2$ | 78,941 | 78,947 | | |
| | $C_3$ | 98,617 | 98,622 | | |
| B3 | $C_1$ | 0.0149 | 0.0149 | 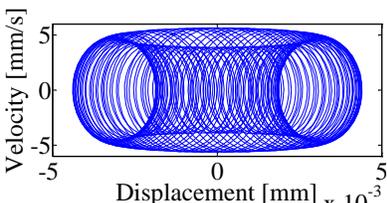 | 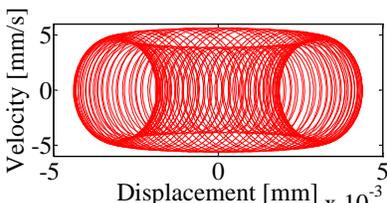 |
| | $C_2$ | 78,941 | 78,947 | | |
| | $C_3$ | 25,952 | 25,958 | | |

According to Table 1, the comparison between the coefficients of the equation of motion obtained from the theoretical model and those derived using the SINDy algorithm reveals a maximum difference of 0.04% for Benchmark (B) 1, 0.04% for B2, and 0.05% for B3. These studies indicates that SINDy is capable of accurately capturing and predicting the dynamics of the system based on the theoretical model. Table 1 shows that by constructing a library of candidate functions based on knowledge from the theoretical model, we ensure high accuracy [24-26]. For theoretical evaluation SINDy algorithm resilience to noise see Supplementary material, S4.

**Application of SINDy to acoustically large object**

We use time series data of three different experimental measurements to approximate the nonlinear equation of motion. Similar to the previous section we chose amplitudes, here $\mathbf{A_{in}} = [0.45, 0.60, 0.70]$ mm, and but only a single excitation frequency of $\omega_o = 32$ Hz and extracted 4 time series using the experimental setup depicted in Fig. 1. The following equations of motion are obtained after applying SINDy and averaging the extracted coefficients, with $x$ representing the object's displacement from its static equilibrium point and $\dot{x}$ being the object's velocity:

$$\begin{aligned}
\begin{matrix} A_{in,1} \\ A_{in,2} \\ A_{in,3} \end{matrix} := \begin{bmatrix} 1 \\ 1 \\ 1 \end{bmatrix} \ddot{x} + \begin{bmatrix} 160{,}985 \\ 85{,}680 \\ 98{,}467 \end{bmatrix} x - \begin{bmatrix} 351.1 \\ 1{,}212.1 \\ 396.2 \end{bmatrix} \dot{x} - \begin{bmatrix} 667{,}030 \\ 358{,}700 \\ 565{,}600 \end{bmatrix} x^3 + \begin{bmatrix} 10{,}283 \\ 10{,}549 \\ 2{,}566 \end{bmatrix} x^2 \dot{x} \\
- \begin{bmatrix} 8.64 \\ 19.02 \\ 7.17 \end{bmatrix} x \dot{x}^2 + \begin{bmatrix} 0.2903 \\ 0.2470 \\ 0.0688 \end{bmatrix} \dot{x}^3 - \begin{bmatrix} 1{,}884 \\ 889 \\ 439 \end{bmatrix} x^3 \dot{x} - \begin{bmatrix} 0.4985 \\ 5.2717 \\ 1.8023 \end{bmatrix} x^2 \dot{x}^2 \\
+ \begin{bmatrix} 398{,}500 \\ 188{,}530 \\ 1{,}452{,}000 \end{bmatrix} x^5 + \begin{bmatrix} 60.452 \\ 170.34 \\ 34.93 \end{bmatrix} x^3 \dot{x}^2 - \begin{bmatrix} 4.6975 \\ 1.6891 \\ 0.4018 \end{bmatrix} x^2 \dot{x}^3 \\
- \begin{bmatrix} 9255 \\ 14{,}647 \\ 17{,}641 \end{bmatrix} x \sin(\omega t) + \begin{bmatrix} 9{,}254 \\ 14{,}636 \\ 17{,}648 \end{bmatrix} \sin(x) \sin(\omega t) = 0
\end{aligned} \quad (7)$$

The same extended function library, including polynomial terms in $x$ and $\dot{x}$ up to fifth order, was used for both the small and large object cases. The differences in the final identified models are a result of the SINDy algorithm's sparsity-promoting structure. For the small object, most coefficients were eliminated, indicating relatively simpler dynamics. For the large object, more terms were retained which represent its richer nonlinear behavior. This consistent approach avoids bias and allows the results to reflect the underlying system response. To generate a formulation using SINDy that not only displays the average of the measured data but also includes uncertainty bounds, we can use the mean value $\pm 2\sigma$ for each coefficient, which covers 95% of the data for a normal distribution. This approach provides a more conservative and comprehensive representation of uncertainty (see Tables S1-S3, in Supplementary material, S5). Fig. 2 presents a statistical analysis of the variation in the calculated coefficients, using box plots [42-46] to illustrate the distribution of the SINDy-obtained coefficients for reconstructing the equation of motion based on the performed measurements.

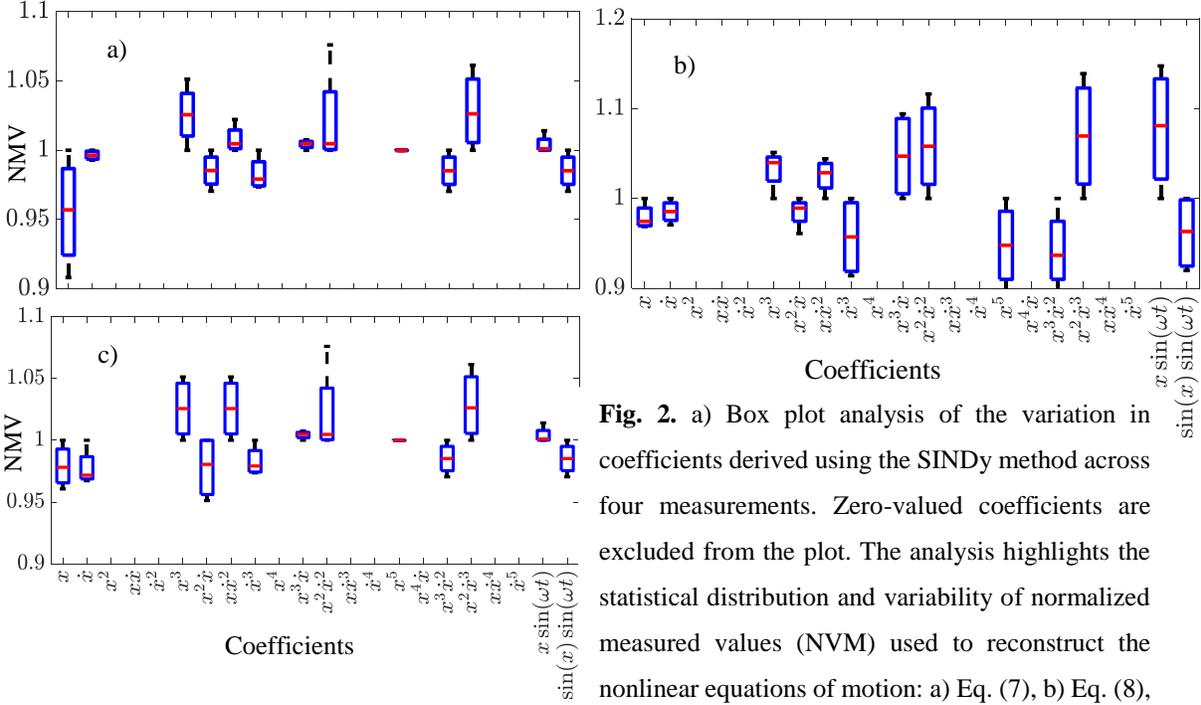

**Fig. 2.** a) Box plot analysis of the variation in coefficients derived using the SINDy method across four measurements. Zero-valued coefficients are excluded from the plot. The analysis highlights the statistical distribution and variability of normalized measured values (NVM) used to reconstruct the nonlinear equations of motion: a) Eq. (7), b) Eq. (8), and c) Eq. (9)).

To ensure comparability, all coefficients were normalized by dividing each by its respective maximum value. To view the original values of the coefficients used to reconstruct Eq. (7), refer to the coefficients listed in Tables S1, S2, and S3 in the Supplementary Material. Zero-valued coefficients are excluded from the plot to focus on the relevant dynamics. When a coefficient shows a small box plot in one subfigure and a large box plot in another—corresponding to different excitation amplitudes—it indicates that the coefficient's variability is sensitive to the excitation amplitude. The current uncertainty assessment, as depicted in Fig. 2, relies on statistical variation across repeated experimental runs and coefficient distributions. However, in the literature, there are other approaches to uncertainty quantification in sparse identification. One such approach, UQ-SINDy, is gaining traction [47], which integrates probabilistic modeling into the identification process by estimating distributions over both the model structure and coefficients [47]. UQ-SINDy has the potential to yield confidence intervals directly on terms within the governing equation and support model selection under uncertainty. Although beyond the scope of this study, incorporating uncertainty-aware methods such as UQ-SINDy could enhance model reliability in future work, particularly under noisy or ambiguous experimental conditions.

When the governing equation of motion, Eq. (7), changes only in its coefficients due to variations in the bifurcation parameter, $A_{in}$, we can describe this as a parametric variation within the same dynamical model. Fig. 3 compares the measured experimental data, the filtered time series, and the SINDy results, along with the power spectral density (PSD) of the LDV measured signal before and after filtering, as well as the PSD according to the SINDy results in Eq. (7) for three different measurements by changing the amplitude of the external excitation, $\mathbf{A_{in}}$, respectively.

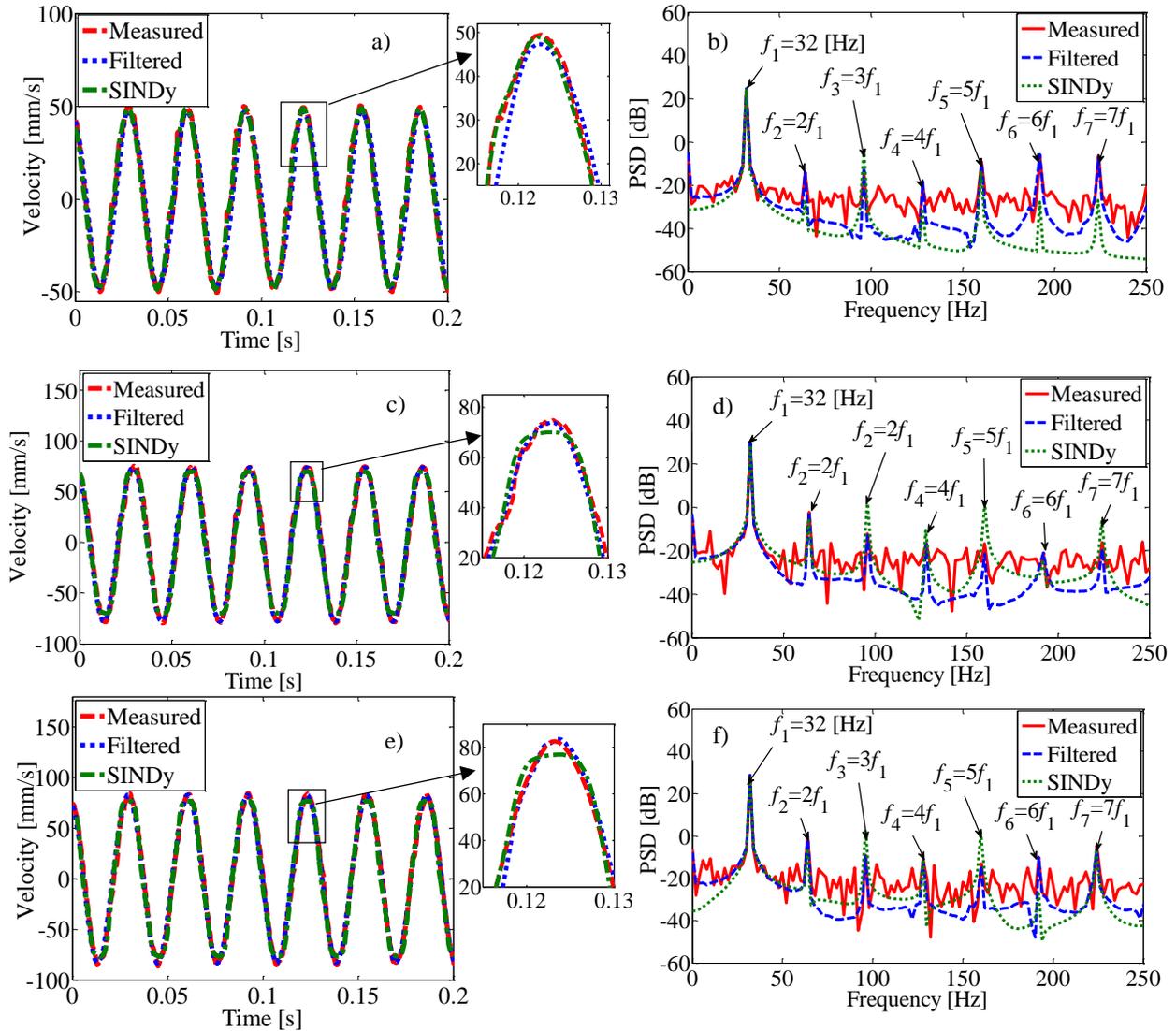

**Fig. 3.** Comparison of experimentally measured data and SINDy predictions using averaged coefficients with filtered data and PSD of the original signal measured by LDV before applying a filter, after applying the filter for: a, b) Measurement 1, c, d) Measurement 2, e, f) Measurement 3. These plots highlight the accuracy of the SINDy algorithm in predicting the system dynamics

The time series presented in Fig. 3(a–f) correspond to steady-state oscillations of the levitated object, recorded after initial transients had subsided under continuous acoustic excitation. These intervals were deliberately selected for modeling, as they exhibit consistent and repeatable dynamics representative of the system's nonlinear behavior. Transient segments—such as those arising during start-up or excitation changes—were excluded to avoid variability that does not reflect the system's long-term response. By focusing on well-defined attractor states, the identified governing equations capture the essential features of the sustained dynamics and remain robust within the experimentally accessible range.

In Fig. 3, the alignment of the PSD peaks from the measured signal, filtered signal and SINDy-reconstructed data suggests that SINDy successfully captured the essential dynamics of the system. It shows the fundamental frequency of 32 Hz and its integer multiples of a fundamental frequency, and its harmonics $2f, 3f, 4f$, indicative for a nonlinear period-1 limit cycle being present [41,42]. The PSD further confirms that the approximation provided is quite accurate. Since SINDy relies on a predefined library of candidate functions whose suitability affects the model's accuracy the library we had chosen has been appropriate for the problem at hand [24-26].

To further validate the results, we next apply the Gorkov formulation and expand the Taylor series to the fifth order to study small objects levitated in air subjected to external excitation (see Supplementary Material S2). This test will provide insides, whether the equations extracted from SINDy will be structurally different to the original Gorkov formulation. The resulting governing equation of motion is therefore derived as follows:

$$\ddot{\theta} + C_1\theta + C_2\theta^3 + C_3\theta^5 + C_d\dot{\theta}|\dot{\theta}| = F\cos(\theta)\sin(\omega t), \tag{8}$$

where the coefficients $C_i$ with $i = 1,2,3$ related to the object and fluid properties and $F$ represents the amplitude of the external excitation. By using SINDy on a large spherical object, Eq. (8) takes the following form

$$\ddot{\theta} + c_1\theta + c_2\theta^3 + c_3\theta^5 + \mathcal{F}(\theta, \dot{\theta}) = (f_1\theta + f_2\sin(\theta))\sin(\omega t), \tag{9}$$

when $\mathcal{F}(\theta, \dot{\theta})$ is a nonlinear function and can be written as

$$\mathcal{F}(\theta,\dot{\theta}) = d_1\dot{\theta} + d_2\theta^2\dot{\theta} + d_3\theta\dot{\theta}^2 + d_4\dot{\theta}^3 + d_5\theta^3\dot{\theta} + d_6\theta^2\dot{\theta}^2 + d_7\theta^3\dot{\theta}^2 + d_8\theta^2\dot{\theta}^3, \quad (10)$$

in which all $c_i$ coefficient in Eq. (9) and $d_i$ coefficient in Eq. (10) are related to the external excitation amplitude, and $\mathcal{F}(\theta,\dot{\theta})$ is a nonlinear function of $\theta,\dot{\theta}$ that can be replaced instead of $C_d\dot{\theta}|\dot{\theta}|$ in Eq. (10). Also, the right hand of the Eq. (10), $F\cos(\theta)\sin(\omega t)$, has be replaced by $(f_1\theta + f_2\sin(\theta))\sin(\omega t)$. All coefficients in Eq. (9) and (12) can be obtained using the SINDy algorithm. From measurements one to three one can see that $f_1$ and $f_2$ are very close to each other and can be treated as a single coefficient, $f$. The ratio of $f$ to $A_{in}$ in all three equations is about 25,000. This suggests a consistent relationship between $f$ and $A_{in}$, which might be an important characteristic of the system. In comparison with Eq. (8), Eq. (9) indicates a stronger influence of viscosity and increasingly important interaction-effects between object, fluid, and the system's dynamic response. The SINDy method reveals important nonlinear effects in acoustically levitated objects that classical models like Gorkov's miss. Specifically, cubic and mixed nonlinear terms show complex interactions between acoustic forces, drag, and inertia at larger particle sizes. These nonlinear terms depend on excitation amplitude, meaning the trap's stiffness and damping change with vibration strength. Also, strong velocity-dependent terms for large objects indicate that energy loss grows at higher excitations—something hard to capture with traditional analysis. This new understanding of scaling and nonlinear behavior can help improve control and characterization in future levitation systems.

Fig. 4a presents the bifurcation diagram for an acoustically large object at $k = 1$ and r = 1.87 mm, where the Gorkov formulation (red line) fails to predict the experimental system behavior (blue dots). Fig. 4b uses instead of Gorkov the extracted equation from SINDy (red line) as analytical result, with experimental data showing up again as blue dots which shows a much better match.

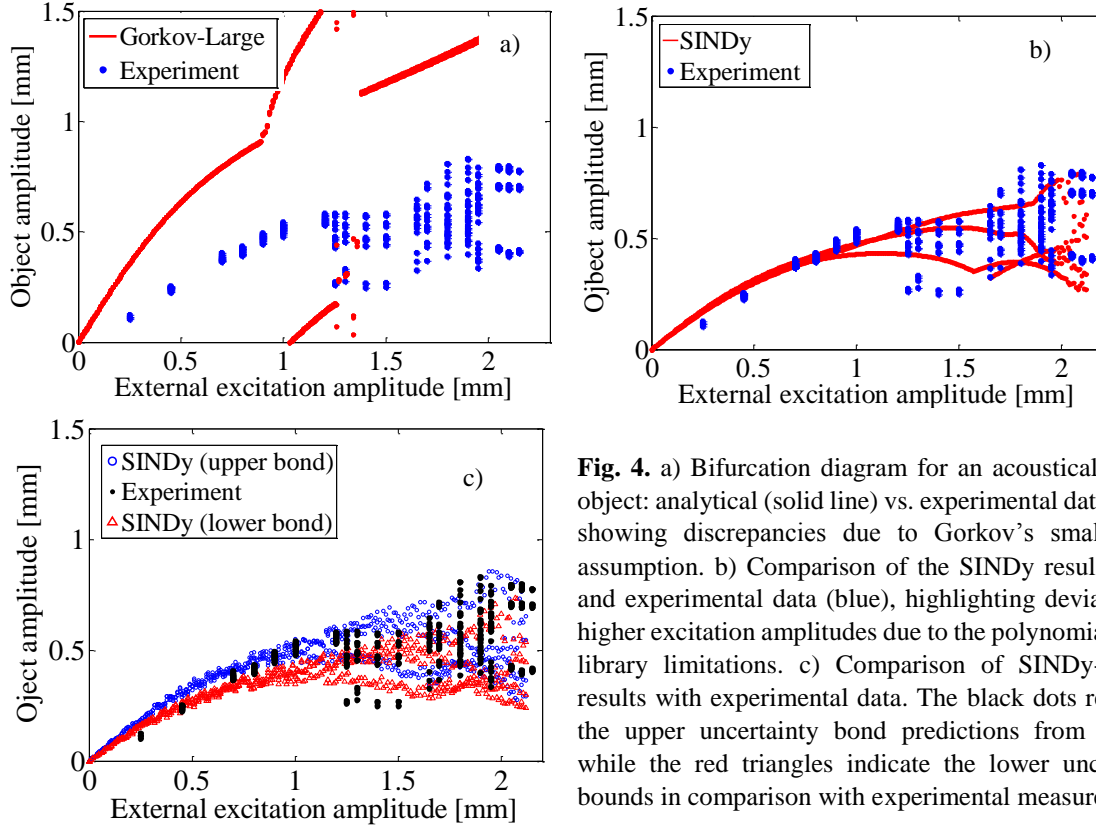

**Fig. 4.** a) Bifurcation diagram for an acoustically large object: analytical (solid line) vs. experimental data (dots), showing discrepancies due to Gorkov's small-object assumption. b) Comparison of the SINDy results (red), and experimental data (blue), highlighting deviations at higher excitation amplitudes due to the polynomial-based library limitations. c) Comparison of SINDy-derived results with experimental data. The black dots represent the upper uncertainty bond predictions from SINDy, while the red triangles indicate the lower uncertainty bounds in comparison with experimental measurements.

Fig. 4a highlights the divergence between the theoretical bifurcation structure and the experimental results at higher excitation amplitudes. Specifically, Gorkov's model fails to capture the multi-branch bifurcation patterns observed in the experiment, indicating its inability to represent the system's nonlinear transitions and amplitude-dependent behavior. In contrast, Fig. 4b demonstrates that the SINDy-derived model captures both the fundamental and higher-order nonlinearities, providing significantly better agreement with the experimental bifurcation data. The results obtained using the statistical analysis are presented in Fig. 4c. The red markers represent the lower bond predicted by the SINDy model, while the black dots illustrate the statistical upper uncertainty bounds also the experimental data are also included for comparison. The bifurcation diagrams in Fig. 4 reflect how the system responds to changes in the excitation amplitude. Although we do not derive a closed-form normal form using analytical methods such as center-manifold reduction, the SINDy-identified equations can be considered a data-driven normal form, where the external excitation amplitude acts as a bifurcation parameter. These results suggest that SINDy provides not only an interpretable model but also a practical tool for reconstructing nonlinear bifurcation behavior in experimentally limited regimes.

## Conclusion

In this study, we applied the Sparse Identification of Nonlinear Dynamical Systems (SINDy) algorithm to reconstruct a levitated object's nonlinear differential equation of motion trapped in an acoustic radiation force field, using theoretical and experimental time series. To evaluate the accuracy and reliability of the SINDy framework, we performed a two-stage validation approach. First, we conducted theoretical validation by applying SINDy to time series data generated from a known nonlinear model (Eq. 1) using a fourth-order Runge-Kutta method. We found that the extracted coefficients differed by less than 0.05% from the exact analytical model. This demonstrates that SINDy is capable of accurately recovering the underlying dynamics of a known nonlinear oscillator. In the second stage, we applied SINDy to experimental data from acoustically large objects. The experimental data used in this study correspond to the system's response under a period-1 oscillatory regime, which was the only stable and repeatable dynamic state achievable within the limits of our acoustic levitation setup. While this restricts experimental validation to a single regime, it still provides a rich basis for extracting governing dynamics. Importantly, the structure of the identified model is not hand-tuned but emerges purely from the SINDy algorithm's sparse regression process, using the same candidate function library across both small and large object cases. This allows for a direct comparison: in the large-object regime, SINDy revealed significant higher-order and mixed nonlinear terms that were absent in the small-object case, highlighting how the model captures physically meaningful transitions beyond the assumptions of classical formulations. These findings demonstrate that, even within a limited dynamic regime, the identified equations carry clear physical signatures and offer insights into amplitude-dependent dynamics not previously characterized in the literature.

We used box plots to visualize distributions and to assist in identifying key coefficients in SINDy models, to reveal their sensitivity to changes in external excitation. For acoustically large objects, SINDy reveals nonlinear dependencies between the external excitation amplitude and the extracted coefficients in the governing equation. These relationships indicate that the system's response is strongly amplitude-dependent. Although the chosen library may not be used to fully capture all aspects of the bifurcation behavior, it provides valuable insights into the dominant dynamics of the system. By combining rigorous theoretical modeling with robust experimental validation, this study provides a data-driven framework for understanding and predicting the dynamics of acoustically manipulated objects. Such a framework could be extended to applications in biomedical engineering—for example, in non-contact manipulation of drug carriers or cells using focused ultrasound. In precision manufacturing, it could inform the stable levitation and positioning of components or objects without mechanical contact. Similarly, materials science may support the characterization of soft or lightweight materials through dynamic levitation-based testing methods.

## Author contributions

Author contributions have been evaluated according to CRediT.

Mehdi Akbarzadeh: Conceptualisation, Methodology, Software, Validation, Formal Analysis, Investigation, Data Curation, Writing - Original Draft, Writing – Review and Editing, Visualisation. Benjamin Halkon: Investigation, Writing - Review and Editing, Supervision. Sebastian Oberst: Conceptualisation, Methodology, Investigation, Validation, Resources, Writing - Review and Editing, Supervision, Project administration, Funding acquisition.

## Acknowledgements

The authors would like to thank Dr. Shahrokh Sepehrirahnama for some early discussions with the lead author.

## Funding Declaration

The first author acknowledges a UTS President's Scholarship. This research is partially supported by the Australian Research Council Discovery Project DP240101536.

## Data availability

The datasets generated and analyzed during the current study are available in the Zenodo repository: https://zenodo.org/records/16033764.

The data include nonlinear time series measurements from the acoustic levitation experiments described in the study, "Sparse identification of nonlinear dynamics applied to the levitation of acoustically large objects." While the full dataset is provided, only the first three Period-1 oscillation cycles were used in the analysis presented in the paper. The accompanying MATLAB code for data processing and system identification was written and tested using MATLAB R2023a. For further inquiries, please contact the corresponding author.